\documentstyle[preprint,pre,aps,tighten]{revtex}
\tighten
\begin{document}
\preprint{}
\draft
%
%%%%%%%%%%%%%%%%%%%%%%%%%%%%%%%%% TITLE PAGE
%
\title{\normalsize \bf
AMBIPOLAR TUNNELING IN NEAR-SURFACE QUANTUM WELLS}
\author{V. Emiliani, A. Frova, and C. Presilla \\
Dipartimento di Fisica, Universit\`a di Roma ``La Sapienza'',\\
Piazzale A. Moro 2, Roma, Italy 00185 }
%\address{Dipartimento di Fisica, 
%Universit\`a di Roma ``La Sapienza'',\\
%Piazzale A. Moro 2, Roma, Italy 00185 \\
%}
\date{Superlattices and Microstructures {\bf 20} (1996) 1-6}
\maketitle
%
%%%%%%%%%%%%%%%%%%%%%%%%%%%%%%%%% ABSTRACT
%
\begin{abstract}
\normalsize        
We study the photoluminescence from a near-surface quantum well
in the regime of ambipolar tunneling to the surface states.
Under steady-state excitation an electric field develops 
self-consistently due to the condition of equal  
tunneling currents for electrons and holes. 
The field induces a Stark shift of the photoluminescence  signal
which compares well with 
experimental data from near-surface GaAs/AlGaAs single quantum wells.
								      
\end{abstract}
%
%%%%%%%%%%%%%%%%%%%%%%%%%%%%%%%%% PACS NUMBERS
%
%
%%%%%%%%%%%%%%%%%%%%%%%%%%%%%%%%% PAPER BODY
%
\vfill
\newpage

For a quantum well built in proximity of an unpassivated surface,
tunneling to surface states can be a nonradiative recombination channel 
competitive with photoluminescence.
The importance of this effect in determining the emission efficiency
has been demonstrated experimentally in various papers \cite{1,2,3,4}.
Recently we have proposed a quantitative model based on ambipolar 
tunneling of electrons and holes which is applicable to many-well
systems in the bulk or single wells coupled to surface states \cite{5}.
In steady-state situations the ambipolar regime, with equal tunneling 
currents for electrons and holes, imposes an electric field
to develop \cite{6,7,8,9}. 
The field induces a peak shift of the excitonic 
recombination via the quantum confined Stark effect \cite{10}.
Here we specialize the discussion to the case of a
quantum well coupled to surface states and we compare the 
theoretical results with photoluminescence experimental data
in GaAs/AlGaAs material.

We will use the following notation. 
The width of the quantum well is $a$ and the width of the 
surface barrier is $b$. 
The bottom of the $e1$ and $hh1$ bands of the well are 
$E_{e1}$ and $E_{hh1}$
and $G$ is the generation current density of electron-hole pairs in the well.
We assume that no pairs are generated within the barrier or at the surface.
The pairs generated into the well relax almost instantaneously,
compared to the other relevant time scales, to the lowest band of the well. 
Electron-hole interaction leads to exciton formation.
Tunneling from the well to the surface states is due 
to free electrons and holes only \cite{6}. 
On the other hand, photoluminescence is restricted only to excitonic 
recombination in the well.
If $n_w$ and $p_w$ are the steady-state concentrations 
(number of particles per unit area) of electrons and holes 
in the well,  the following rate equations hold:
\begin{mathletters}
\label{RE}
\begin{equation}
0 = G - J_{e} - \lambda n_w p_w
\end{equation}
\begin{equation}
0 = G - J_{h} - \lambda n_w p_w
\end{equation}
\begin{equation}
0 = \lambda n_w p_w  - I  ~.
\end{equation}
\end{mathletters}

The bimolecular generation rate of excitons is assumed  
proportional to the electron and hole concentrations \cite{11}.
The photoluminescence current density $I$ is proportional to the exciton 
concentration in the well. 
Transfer of electrons (holes) from the well to the surface states 
is realized in a non-coherent two-step process. 
Quantum coherent tunneling of electrons (holes) from an occupied state of
the $e1$ ($hh1$) band of the well to an equal-energy empty state at 
the surface is followed by relaxation toward the lower energy states. 
When the barrier width $b$ is not too small, 
the phonon relaxation process at the surface is much faster than
the tunneling process (current densities $J_e$ and $J_h$) 
and can be neglected. 

The tunneling current densities are approximately proportional to the charge 
concentrations in the well and the proportionality factor, namely
the tunneling probability, is generally quite different for electrons and holes.
Therefore, in a steady-state situation 
when $J_{e}=J_{h}$, the concentrations of electrons and holes
in the well must be different.
The resulting electric field, in turn, affects the electron and hole 
tunneling probabilities.

Since the tunneling rate depends both on the effective mass of the
carriers and the density of states at the surface, two cases are
possible.
When the density of states of the donor-like band at the surface is not
sufficiently smaller than that of the acceptor-like band, 
the electron tunneling rate is larger than the hole tunneling rate.
In this case electrons accumulate at the surface and in a steady-state 
situation $p_w>n_w$.
The electric field is directed from the well to the surface and its value
is given by
\begin{equation}
F={en \over \varepsilon_0 \varepsilon_r} 
\label{F}
\end{equation}
where $n=p_w-n_w$ and $\varepsilon_r$ is the permittivity
of the barrier material. 
A reversed situation, however, may happen when the effective mass
difference between electrons and holes is overcompensated by
the difference in the surface densities.

For a given value of the electric field $F$, 
in the first order perturbation theory 
the tunneling current densities are
$J_{e}=n_w/\tau_e$ and $J_{h}=p_w/\tau_h$ where
\begin{equation}
{1 \over \tau_e} = {2 \pi \over \hbar} 
\left| \langle \Phi^{s}_0 |V_e| \Phi^{w}_0 \rangle \right|^2  
A~ \nu_e(0)
f\left({ \epsilon_F^e \over k_B T} \right) 
\label{TAUE}
\end{equation}
\begin{equation}
{1 \over \tau_h} = {2 \pi \over \hbar} 
\left| \langle \Phi^{s}_0 |V_h| \Phi^{w}_0 \rangle \right|^2  
A ~\nu_h(0)~
f\left({ \epsilon_F^h \over k_B T} \right) ~.
\label{TAUH}
\end{equation}
$\nu_e(\epsilon)$ and $\nu_h(\epsilon)$ are the densities of states
in the donor-like and acceptor-like surface bands respectively.
Energies are measured from the bottom of the $e1$ and $hh1$ bands.
The electron and hole Fermi energies $\epsilon_F^e$ and $\epsilon_F^h$
which appear in the Fermi function $f$
are related to the respective electron and hole concentrations  
at the surface, $n_s$ and $p_s$, as explained in the following.
Finally, $A$ is the relevant transverse area and the matrix elements are 
evaluated in the Appendix.

When the electric field and the tunneling rates are known, the solution
of the rate equations is 
\begin{mathletters}
\label{SORES}
\begin{equation}
n_w = \sqrt{ \left({n\over2}+{1\over2\lambda \tau_e} \right)^2 +
{G \over \lambda} } -
\left({n\over2}+{1\over2\lambda \tau_e} \right)
\end{equation}
\begin{equation}
p_w=n_w+n ~.
\end{equation}
\end{mathletters}
Moreover, $n_s=n$ and $p_s=0$ if $F>0$ and
$n_s=0$ and $p_s=-n$ if $F<0$.

This result allows one to find the steady-state values of the 
electric field, of the charge concentrations and of the tunneling rates 
by a recursive method.
Starting from some trial value,
the electric field, i.e. the charge concentration $n$, is
changed until the condition $J_{e}=J_{h}$ is reached. 
At this point the luminescence current density $I$ is obtained 
from the equilibrium concentrations of electrons and holes in 
the well.

Carrying out explicitly the calculations implies the knowledge 
of the energy distribution of the surface states. 
Inversely we can try to get information on the surface states by fitting 
experimental photoluminescence data.
We concentrate on the specific example of an Al$_{0.3}$Ga$_{0.7}$As 
surface with a nearby GaAs quantum well \cite{1,6}.

At energy close to the bottom of the $e1$ band of the well the  
Al$_{0.3}$Ga$_{0.7}$As surface has only donor-like states
belonging to the exponentially vanishing Urbach tail
\begin{equation}
\nu_e(\epsilon) = { m_e^s \over \pi \hbar^2 } 
\exp \left( -{\Delta E_c + eFb - E_{e1} - \epsilon 
\over \epsilon_e} \right)  ~.
\label{URBACH}
\end{equation}
Note that energy is measured from the bottom of the $e1$ band.
Such states are assumed to be nodal hydrogenic wavefunctions 
\cite{12} with radius $r_e$ fixed by their depth into the gap.
Their explicit expression is given in the Appendix.
We assume that at the top of the gap the state density is the 
two dimensional density of free Al$_{0.3}$Ga$_{0.7}$As electrons 
with effective mass $m_e^s$.
The parameter $\epsilon_e$ will be considered as a fitting parameter.
According to Eq.\ (\ref{URBACH})   
the Fermi energy for the donor-like surface band containing $n_s$ 
electrons is
\begin{equation}
\epsilon_F^e =  \Delta E_c + eFb - E_{e1} + \epsilon_e 
\ln \left( {\pi \hbar^2 n_s \over m_e^s \epsilon_e} \right) ~.
\end{equation}

On the other hand, at energy close to the bottom of the $hh1$ band 
of the well the Al$_{0.3}$Ga$_{0.7}$As surface has a very high 
concentration of acceptor-like defect states \cite{13}. 
We schematize them again by nodal hydrogenic wavefunctions
\cite{12} but with radius $r_h$ to be considered as a second fitting 
parameter.
These states are assumed to be distributed in energy with constant 
density $\nu_h$ over an interval $\Delta E_h$ into the gap.
The Fermi energy for the acceptor-like surface band is then
\begin{equation}
\epsilon_F^h = p_s/\nu_h + \Delta E_v - eFb - E_{hh1} -\Delta E_h ~.
\end{equation}

Due to the high ratio between the hole and the electron surface-state
density \cite{13}
holes accumulate at the surface and in a steady-state situation we have
$F<0$. 

Experimental photoluminescence data are available 
(see Ref. \cite{1} for details) for a 
well width $a=60$ \AA\ , a temperature $T=4.2$ K,
a photon-pump energy $h\nu=1.608$ eV and with an incident power density 
$P_i=0.5$ Wcm$^{-2}$. 
The absorption efficiency is estimated to be 1\%, so we take
$G=0.01~P_i/h\nu$.
The relevant material parameters are \cite{14}:
$\Delta E_c=0.3$ eV, $\Delta E_v=0.128$ eV, 
$m_e^s=0.091~m$, $m_e^w=0.067~m$, $m_h^w=0.34~m$, 
$m$ being the free electron mass, and $\varepsilon_r=12$. 
Moreover we put $\lambda=6$ cm$^2$ s$^{-1}$ \cite{11}.
We assume an acceptor-like surface state density 
$\nu_h=10^{14}$ cm$^{-2}$eV$^{-1}$ \cite{15} with 
$\Delta E_h \simeq 0.5$ eV
(the results we found do not depend crucially on this particular value).

The free parameters, $\epsilon_e$ and $r_h$, are fixed by 
fitting the normalized photoluminescence intensity $I/I_\infty$
to the experimental data \cite{1} obtained for different values of 
the barrier width $b$ (the normalization factor $I_\infty$ is
the photoluminescence current density for $b \to \infty$). 
A least-square-error procedure gives the unique solution  
$\epsilon_e=12$ meV and $r_h=11$ \AA.
In Fig. 1 we compare the ratio $I/I_\infty$, 
calculated with these values, with the experimental data. 
The agreement is excellent.

In Fig. 2 we show the electric field value calculated in the situation 
of Fig. 1 as a function of the barrier thickness.
In the same figure we show also the field obtained with different values
of the incident power density $P_i$, all the other parameters being fixed.
It is seen that, for high levels of excitation, the field approaches 
values of order $10^{5}$ V cm$^{-1}$ and
keeps increasing when the barrier becomes thinner. 

An important check of the validity of our model is given by the
analysis of the Stark shift.
The self-consistently estimated electric field induces a band bending
which modifies the single particle levels $e1$ and $hh1$ and, 
therefore, the exciton recombination energy $\Delta E_p$.
The Stark shift calculated as a sum of the shifts of levels $e1$ and $hh1$
is shown in Fig. 3 as a function of the QW excitation
(details of the calculation will be reported elsewhere).
In Fig. 3 we show also the corresponding measured energy shifts (dots).
The agreement is good in the region ($P_i \simeq 0.5$)
where the fitting parameters were fixed as explained above
and is fairly satisfactory over three orders of magnitude.
An improvement should be possible if the surface-state spectrum were 
known {\em a priori}. 
This is a confirmation that the ambipolar tunneling
approach provides a reasonably accurate description of the 
loss of efficiency in near-surface quantum wells.

%
%%%%%%%%%%%%%%%%%%%%%%%%%%%%%%%%% ACKNOWLEDGMENTS
%
%\acknowledgments

%
%%%%%%%%%%%%%%%%%%%%%%%%%%%%%%%%% APPENDIX
%
%\appendix
\centerline{\bf Appendix}

\centerline{ }

The surface is defined by the plane $z=0$ and the well is in $b<z<b+a$.
Firstly, we consider the case of electrons.
We assume a rectangular potential profile with 
left and right discontinuities 
$V_l=\Delta E_c+eFb/2$ and $V_r=\Delta E_c$ for the well where
$F$ is the electric field in the barrier region $0 \leq z \leq b$.
The electron wavefunction at energy $\epsilon=0$ from the bottom
of the $e1$ band of the well is given by 
\begin{equation}
\Phi^{w}_{\epsilon=0} =
{C \over \sqrt{A} }  
\left\{  
\begin{array}{lll}
\sin\delta \exp \left (k_1 [z-b] \right)             & \mbox{$z<b$}   \\
\sin(k (z-b)+\delta )                
& \!\!\!\!\!\!\!\!\!\!\mbox{$b<z<b+a$} \\
\sin(k a+\delta ) \exp \left (-k_0 [z-b-a] \right) & \mbox{$z>b+a$}
\end{array}  
\right.
\label{PHI2}
\end{equation}
where
$\hbar k_1 = \sqrt{{2m_e^w}(\Delta E_c + eFb/2 - E_{e1})}$, 
$\hbar k = \sqrt{{2m_e^w} E_{e1} }$ and
$\hbar k_0 = \sqrt{{2m_e^w}(\Delta E_c - E_{e1} )}$.
The phase shift is $\delta=\tan^{-1}(k/k_1)$ and the energy
is determined by solving
\begin{equation}
k a = \pi -
\sin^{-1} \left( {\hbar k \over \sqrt{2m_e^{w} V_l}} \right) -
\sin^{-1} \left( {\hbar k \over \sqrt{2m_e^{w} V_r}} \right) 
\end{equation}
The constant $C$ is fixed by normalization
\begin{eqnarray}
C= && \left\{ {\sin^2 \delta \over 2 k_1} +
{\sin^2 (k a + \delta) \over 2 k_0} +
{a \over 2} \right. \nonumber \\
&& \left. ~~~~~~~~~~~~~~~ - {\sin(2(k a + \delta)) - \sin (2 \delta)
\over 4 k} \right\} ^{-1/2}
\end{eqnarray}

The donor-like surface state $\Phi^s_{\epsilon=0}$ is approximated by 
a truncated $2p$ hydrogenic wavefunction \cite{12}  
\begin{equation}
\Phi^s_{\epsilon=0} =
{z \over 4 \sqrt{\pi} r_e^{5/2} } \exp \left ( r/2 r_e \right)
\left\{  
\begin{array}{ll}
0  ~~~   & \mbox{$z<0$} \\
1  ~~~   & \mbox{$0<z$} 
\end{array}  
\right.
\label{PHI1S}
\end{equation}
where $r=\sqrt{x^2+y^2+z^2}$.
The state is at energy $\hbar^2 / (8 m_e^s r_e^2)$ below
the bottom of the conduction band for the barrier material where 
the electron effective  mass is $m_e^s$.
By imposing the condition that this energy corresponds to $\epsilon =0$,
we determine the
radius  $r_e$ 
\begin{equation}
r_e = {\hbar \over \sqrt{ 8 m_e^s 
\left( \Delta E_c - E_{e1} + eFb  \right) } }
\label{RAE}
\end{equation} 

Assuming that the perturbation potential $V_e$ is of the order 
of the conduction band offset,
the tunneling matrix element between the  
well and surface states at $\epsilon=0$ can be evaluated analytically
\begin{eqnarray}
& & \langle \Phi^s_0 |V_e| \Phi^w_0 \rangle =
{\sqrt{\pi} r_e^{3/2} \Delta E_c C \sin\delta 
\over 8 \sqrt{A}}  \left\{
e^{-k_1 b} \left[ {1\over 4 (2k_1 r_e-1)^2}
\right. \right. \nonumber \\ 
& & - \left. \left. {1\over 32 (2k_1 r_e-1)^3} \right] \right. + 
e^{-b/2r_e} \left. \left[ {(b/r_e)^2 +2 b/r_e \over 2k_1 r_e-1}
\right. \right. \nonumber \\ 
& & - \left. \left. {1+b/r_e \over 4 (2k_1 r_e-1)^2}+ 
{1\over 32 (2k_1 r_e-1)^3} \right]  \right\}
\label{MEL}
\end{eqnarray}

In the case of holes we have a completely analogous situation 
where the relevant band in the well is $hh1$ instead of $e1$
and the acceptor-like surface state is given by Eq.\ (\ref{PHI1S})
with $r_e \to r_h$.
Equation (\ref{MEL}) gives the tunneling matrix element for holes 
with the substitutions $r_e \to r_h$, $\Delta E_c \to \Delta E_v$, 
$eFb \to -eFb$.

%
%%%%%%%%%%%%%%%%%%%%%%%%%%%%%%%%% REFERENCES LIST
%

%
%%%%%%%%%%%%%%%%%%%%%%%%%%%%%%%%% FIGURE CAPTION
%
\begin{figure}
\caption{Normalized photoluminescence 
ratio $I/I_\infty$ of a near-surface well {\it vs} the surface-barrier 
thickness $b$. Dots: experimental data from Ref. [1]; solid line: 
best fitting in terms of the self-consistent model. Incident power 
density is $P_i$ = 0.5 W cm$^{-2}$.}
\label{FIG1}
\end{figure}

\begin{figure}
\caption{Calculated  electric field $F$
across the surface barrier {\it vs} the 
surface barrier thickness $b$ for different incident power densities $P_i$.}
\label{FI2}
\end{figure}

\begin{figure}
\caption{Comparison between the Stark shift of the photoluminescence 
signal calculated from the model (solid line) and measured (dots)
{\it vs} the incident power density $P_i$. The surface-barrier thickness 
is b = 80\AA. }
\label{FIG3}
\end{figure}   
%
%%%%%%%%%%%%%%%%%%%%%%%%%%%%%%%%% END
%
\end{document}